\documentclass[aoas,MSNbibl,nameyear,seceqn,dvips]{arximspdf}
\usepackage{dcolumn}
\usepackage{graphicx}

\doi{10.1214/12-AOAS566} 
\volume{6}
\issue{4}
\pubyear{2012}
\firstpage{1615}
\lastpage{1640}

\makeatletter

\newcolumntype{d}[1]{D{.}{.}{#1}}

\newcommand{\rrvert}{\vert}
\newcommand{\llvert}{\vert}

\newcommand{\cal}{\mathcal}

\makeatother

\begin{document}
\begin{frontmatter}

\title{Dating medieval English charters\thanksref{T1}}
\runtitle{Dating medieval English charters}

\thankstext{T1}{Supported by a grant from the
Natural Sciences and Engineering Research Council of Canada
and by a grant from Google Inc.}

\begin{aug}
\author[A]{\fnms{Gelila} \snm{Tilahun}\ead[label=e1]{gelila@utstat.toronto.edu}},
\author[A]{\fnms{Andrey} \snm{Feuerverger}\corref{}\ead[label=e2]{andrey@utstat.toronto.edu}}
\and
\author[B]{\fnms{Michael} \snm{Gervers}\ead[label=e3]{m.gervers@utoronto.ca}}
\runauthor{G. Tilahun, A. Feuerverger and M. Gervers}
\affiliation{University of Toronto}
\address[A]{G. Tilahun\\
A. Feuerverger \\
Department of Statistics \\
University of Toronto\\
100 St. George Street\\
Toronto, Ontario M5S 3G3\\
Canada \\
\printead{e1}\\
\hphantom{E-mail: }\printead*{e2}} 
\address[B]{M. Gervers \\
Department of History \\
University of Toronto \\
100 St. George Street\\
Toronto, Ontario M5S 3G3\\
Canada\\
\printead{e3}}
\end{aug}

\received{\smonth{11} \syear{2011}}
\revised{\smonth{4} \syear{2012}}

%
\begin{abstract}
Deeds, or charters, dealing with property rights, provide a continuous
documentation which can be used by historians to study the evolution of
social, economic and political changes. This study is concerned with
charters (written in Latin) dating from the tenth through early
fourteenth centuries in England. Of these, at least one million were
left undated, largely due to administrative changes introduced by
William the Conqueror in 1066. Correctly dating such charters is of
vital importance in the study of English medieval history. This paper
is concerned with computer-automated statistical methods for dating
such document collections, with the goal of reducing the considerable
efforts required to date them manually and of improving the accuracy of
assigned dates. Proposed methods are based on such data as the
variation over time of word and phrase usage, and on measures of
distance between documents. The extensive (and dated) Documents of
Early England Data Set (DEEDS) maintained at the University of Toronto
was used for this purpose.
\end{abstract}

%

\end{frontmatter}

\section{Introduction}

Our object in this paper is to contribute toward
the development of statistical procedures
for computerized calendaring (i.e., dating)
of text-based documents arising, for example,
in collections of historical or other materials.
The particular data set which motivated this study is the
Documents of Early England Data Set (DEEDS) maintained at the
Centre for Medieval Studies of the University of Toronto.
This data set consists of
charters,
that is, documents evidencing the transfer and/or
possession of land and/or movable property,
and the rights which govern them.
The documents in question date from the tenth
through early fourteenth centuries
and are written in Latin, the administrative language of their time.
They were mostly obtained from cartularies and charter collections
produced in England and Wales, with a few from Scotland.

A peculiarity of that era is that most of the charters that were issued
do not bear a date or other chronological marker.
This is particularly so
from the time of the Conquest in 1066, until about 1307,
when fewer than 10\% of the more than one million surviving charters
bore dates.
(A more complete background to these circumstances
is provided in Section \ref{SectionDescription}.)
Charters dating from the twelfth and thirteenth centuries,
however, are a vital source
for the study of English social, economic and political history,
and significant historical information can be derived
when such charters can be dated or sequenced accurately.
(For some examples, see Section \ref{SectionDescription}.)
The charters comprising the DEEDS data set are
derived from among those charters
which can in fact be accurately dated,
and, specifically, to within a year of their actual issue.
A key aim of the DEEDS project was to produce a reliable data base
from which methods for dating the undated charters could be devised.

The DEEDS data set currently consists of some
10,000 documents, in computer readable form,
taken from published editions of charter sources.
These have all been dated by historians on the basis of
internal dates or other internal chronological markers
such as person or place names, or reference to a datable event.
(Note, however, that
dating manually, for instance, by comparing names,
is prone to errors which can multiply
when charters are used to date other charters;
not infrequent names such as
``William son of Richard son of William son of Richard''
can easily be generationally misaligned.)
One key idea underlying our work is that changes
in language use across time can be used to help
identify the date of an undated document.
For example, a study of dated charters shows that the phrase
``\textit{amicorum meorum vivorum et mortuorum}''
(``of my friends living and dead'')
was in currency between the years 1150 and 1240.
As another example, the phrase
``\textit{Francis et Anglicis}''
(a form of address: ``to French and English'')
was phased out when Normandy was lost by England to the French in 1204.
By combining evidence from many words and phrases,
and/or by examining measures of distance between documents,
our goal is to develop algorithms
to help automate the process
of estimating the dates of undated charters
through purely computational means.

In Section \ref{SectionDescription} we provide
further historical background concerning the charters with which
the DEEDS data set is concerned.
We explain there how it happened that
so many charters had been left undated,
and indicate the importance that dating charters correctly
has for research into the social, economic and political
history of England in the high middle ages.
Following this, we provide a more detailed description
of that part of the DEEDS data set on which our work was based.

In Section \ref{SectionPrevious}
we first briefly discuss some concepts relevant for
statistical processing of text-based documents,
and set down the notation to which we will adhere throughout.
We then review some previous calendaring work
that had been carried out using the DEEDS data set.
In Sections \ref{SectionNearest}, \ref{SectionMaximum}
and~\ref{SectionThird} we discuss
three distinct methods for calendaring undated charters.
The methods described in Section \ref{SectionNearest} are based on
nearest neighbors (kNN);
essentially, these methods average the dates
of documents in a training set which have known dates,
and which are ``closest'' to the one being dated.
This approach requires notions of distance between documents
which we also discuss there,
as well as the selection of tuning parameters using cross-validation.
The method proposed in Section \ref{SectionMaximum}
is based on an analogue of maximum likelihood
which we refer to as the \textit{method of maximum prevalence} (MP).
This method attempts to assign a probability,
at every point in time,
that the document would have randomly been produced then,
and it estimates the date of the document
to be the time at which this probability is greatest.
Finally, in Section \ref{SectionThird}, we propose
a method based on determining the minimum
of a nonparametric quantile regression curve
fitted to a scatterplot of the distances from a
document to be dated to the documents in a test set,
against the known dates of those test documents.
Some asymptotic theory for the estimation methods
is discussed briefly in Section \ref{SectionTheory},
and based on the three statistical methods discussed,
numerical work we carried out using the DEEDS data set
is described in Section~\ref{SectionResults}.
Some concluding remarks are provided in
Section \ref{SectionDiscussion} where avenues for
further work are also indicated.

The method discussed in Section \ref{SectionDescription} is due to R.
Fiallos, but is discussed here in statistical terminology and in
greater detail than in \citet{Fia00}. The methods reviewed in
Section \ref{SectionNearest} are from Feuerverger et~al.
(\citeyear{Feuetal05}, \citeyear{Feuetal08}) and are included here for
comparison and completeness. The maximum prevalence method described in
Section \ref{SectionMaximum} is our main new methodological
contribution. As well, a~key contribution of our work lies in the novel
application of the mentioned estimation methods to historical data of
the type considered here. This work may be seen in the context of other
work in the digital humanities, temporal language modeling and
information retrieval. Some entry points to that literature in the
context of calendaring documents include \citet{deJRodHie05},
Kanhabua and Norvag (\citeyear{KanNor08}, \citeyear{KanNor09}) and the
references therein. For broader context see, for example,
\citet{BerBro05} and \citet{ManRagSch08}.

\section{Description of the data set}
\label{SectionDescription}

The keeping of records pertaining to the ownership
and transfer of property is as old as writing itself,
and dates back to at least
the third millenium BC in Sumeria
where such documents were inscribed on clay.
Consequently, deeds, or charters (as they are known),
provide a continuous
legal documentation which can be used by historians
to study the evolution of social, economic and administrative changes.
For charters to be used in this way, however,
establishing an accurate chronology is important.
Below, we will use the term \textit{charter}
to represent an official legal document, 
often written\vadjust{\goodbreak} or issued by a religious, lay or royal institution,
which typically provides evidence of the transfer
of landed or movable property and the rights which govern them.

It was the fate of England, between the time of the Conquest in 1066
when William the Conqueror (also Duke of Normandy) ascended the English
throne, until the start of the reign of Edward I in 1307, that---in
contrast to the Roman and papal traditions---most charters issued did
not bear a date regardless of the level of society in which the
charters originated. William I introduced into the royal chancery the
then-current Norman custom of issuing charters without dates or other
chronological markers. This custom continued until the reign of
England's sixth post-Conquest (and crusader) king---Richard the
Lionhearted (1189--1199)---when, for the first time, documents issued
from the royal chancery began regularly to include a date. It was,
however, not until the accession of the tenth king, Edward~II
(1307--1327), that the custom of including dates also became universally
adopted by those responsible (ecclesiastics and laymen) for issuing
private charters.

Charters from the twelfth and thirteenth centuries, written in
Latin---the administrative language of the time---are the predominant
source for the study of English social, economic and political history
of that era. It is estimated that at least one million charters have
survived from that nearly 250 year period, some as originals, but most
as copies in cartularies (i.e., deed books). Of these, well over 90
percent do not bear dates, so that fewer than 10\% of them can be dated
at all accurately. Although increasingly less so with the passage of
time, even at the turn of the fourteenth century the percentage of
English charters bearing dates remains modest.

Significant historical information can be derived
when charters can be dated or sequenced correctly
as the following three examples attest:
(i) A~study of donations to the twelfth-century
Order of the Hospital of Saint John of Jerusalem
allowed historians to conclude that the
Order became militarized in response to the
fall of Edessa in 1144,
and to the call for the Second Crusade in 1145.
(ii)~Widespread reluctance to incorporate
the invocation of divine intervention
into legal language of the day
evidences the social unrest in England
under the Papal Interdict of \mbox{1208--1214}.
(iii) With the Crusades came the foundation of the
military-religious orders known as the Templars and the Hospitallers
who financed their activities in part through the management
of properties in Europe and the Middle East.
The relative growth of their estates
in London and its suburbs from the twelfth to the
fourteenth centuries 
confirms without a doubt that as London spread outside
its ancient Roman walls in the twelfth century,
the Templars played a far more significant role
in suburban development, and from a much earlier period,
than did the Hospitallers.
Further background and examples may be found in \citet{Ger00},
\citet{GerHam}, and references therein.

The DEEDS database, maintained at the University of Toronto, is now a
corpus of over 10,000 medieval Latin charters dealing primarily
with\vadjust{\goodbreak}
land and movable objects (grants, leases, agreements, etc.) and rights
regulating their use. The charters in this corpus are all
\textit{dated} ones; they were either dated internally or they
contained sufficient information to enable historians to situate them
to within a year of their issue. These charters were all obtained from
published editions of charter sources covering England and Wales, and a
few from Scotland, and were derived predominantly from the archives of
religious houses and towns, as well as lay institutions such as
colleges and universities. (Note that because the charters were taken
from published sources, they necessarily bear any editorial decisions
made by the publishing author.) The DEEDS project has, as a key
objective, to establish computerized methodologies for dating the vast
number of medieval charters that have not yet been dated in the hope
that, taken together, the dated documents from the database, and those
to which dates can be attributed via statistical and other means, may
allow historians to construct a more precise understanding of the
evolution of English society within that era. We remark that due to the
paucity of surviving documents, and the rarity among them of charters
bearing dates, there is very little in the DEEDS database from before
1160.

Original charters, written on parchment, and bearing the seal
of the issuer or his patron, are rare.
Most of the charters that have survived today
exist as copies in deed books known as cartularies
which were produced periodically during the eleventh
to fifteenth centuries.
(Such copying could occasionally introduce
transcriptional or other changes and inaccuracies.)
Consequently, palaeography and sigillography
generally cannot help in the calendaring process,
leaving the evolution over time of vocabulary usage,
word patterns and document structure as the primary data
from which dating can be carried out.
These charters are preserved today in such repositories
as the National Archives, the British Library,
the archives of Oxford and Cambridge Universities,
in county record offices and in private collections.\vspace*{8pt}

\textit{The data}: Although the DEEDS data set has grown, 3353
documents were available to us when our computations were implemented;
we now describe this data set. Prior to their analyses, certain
preprocessing steps were applied. Dates were mapped into the Julian
calendar. Documents were normalized for variations in spelling, and all
punctuation marks were removed. Names were left unchanged, and just as
they appeared in the document. All numbers appearing in a document were
encased between exclamation signs---thus, \textit{xv} became
!\textit{xv}!---and all numbers were subsequently treated as being the
same distinct word. (We are not referring here to actual dates which
might appear in a document allowing it to be dated without difficulty.)
The determination of distinct words was taken to be case sensitive;
this rule was applied even to the first words of sentences whose first
character was generally in upper case. A sample of a document processed
in this way is provided at the end of this section.

\begin{figure}

\includegraphics{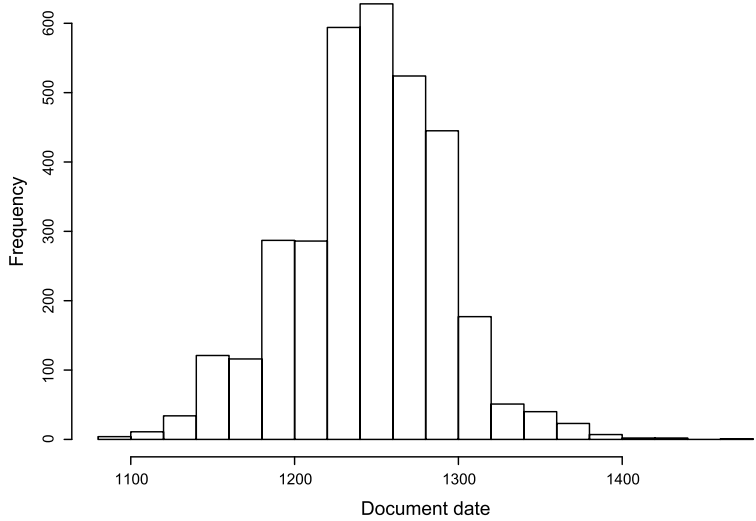}

\caption{Histogram for the
distribution of dates of the 3353 dated documents.}\label{figur1}
\end{figure}

\begin{figure}[b]

\includegraphics{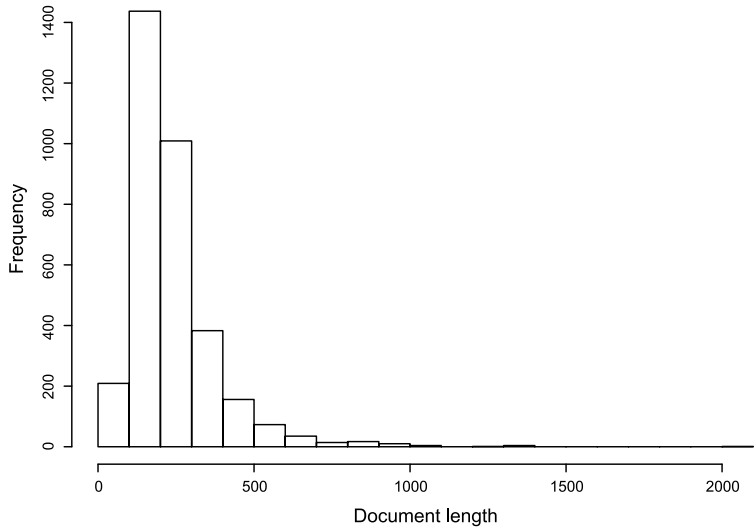}

\caption{Histogram for the
distribution of lengths (word counts) of the 3353 documents.}\label{figur2}
\end{figure}

Figures \ref{figur1} and \ref{figur2}, as well as Table \ref{tabl1}, provide some
graphical and tabular information
%
%
about our 3353 dated DEEDS documents.
Figure \ref{figur1} is a histogram of the known dates for the documents;
the earliest of these is dated 1089,
and the latest is dated 1438.
The mean date of these charters is 1237
with a standard deviation of 46 years.
Figure \ref{figur2} is a histogram of the lengths
(i.e., word counts) of the documents;
the shortest of these consisted of only 15 words,
and the longest of 2054 words;
the median and mean of the word counts were 202 and 237, respectively,
while the lower and upper quartiles were 151 and 275 words.
Very short or very long documents are rare.
Words consisted of an average of 6.5 characters.
No dependencies worthy of note were detected
between the lengths of the documents with their dates,
their contents or with any other features.

Among the 3353 documents, a total of 50,006 distinct words occurred.
Of these, 28,282 words (56\%) occurred only once.
Words which occurred only once were not considered relevant
for our study because such words could not simultaneously occur
in both a test subset and a validation subset of the data.
The frequency of repetition for repeated words
is given in Table~\ref{tabl1}.
While it is possible that in a few instances
such repetitions all occurred within the same document,
we did not keep track of such occurrences.

\begin{table}
\tablewidth=270pt
\caption{Frequency of word repetitions in the
data set of 3353 documents, comprising 50,006 distinct words}\label{tabl1}
\begin{tabular*}{\tablewidth}{@{\extracolsep{\fill}}ld{6.0}@{}}
\hline
\textbf{Word frequency}
& \multicolumn{1}{c@{}}{\textbf{Number of occurrences}} \\
\hline
Words occurring only once & 28\mbox{,}282 \\
Words occurring exactly twice & 7223 \\
Words occurring exactly three times & 3265 \\
Words occurring more than three times & 11\mbox{,}236 \\
Words occurring more than 10 times & 4952 \\
Words occurring more than 30 times & 2330 \\
Words occurring more than 100 times & 1004 \\
Words occurring more than 300 times & 415 \\
Words occurring more than 1000 times & 109 \\
\hline
\end{tabular*}
\end{table}

Finally, we exhibit here one of the DEEDS charters
after preprocessing as indicated above.
This document deals with the transfer of a messuage
(house and appurtenances) in Nottingham
for an annual payment of four pounds sterling.
It bears serial number 00650032 in the DEEDS data set
and has been dated internally by regnal year to 1230--1231:\vspace*{8pt}

\textit{\small
Omnibus sancte matris ecclesie filiis ad quos presens scriptum
pervenerit Simon abbas de Rufford' et conventus eiusdem loci salutem
Noverit universitas vestra nos dedisse concessisse quiete clamasse
et hac presenti carta nostra confirmasse Johanni filio Bele de Notingha'
unum mesuagium cum pertinentiis in Notingha' quod jacet inter terram
Walteri Karkeney et terram Ade de Estweyt habend et tenend
eidem Johanni et heredibus suis et heredibus eorum in feodo et
hereditate de nobis vel atornatis nostris libere quiete integre pacifice
et honorifice reddendo inde annuatim nobis vel atornatis nostris
quatuor solidos sterlingorum ad duos terminos anni scilicet duos solidos
ad Pentecosten et duos solidos ad festum sancti Martini pro omni
servicio consuetudine seculari demanda et exactione Et nos
predictam terram cum pertinentiis predicto Johanni et heredibus suis vel
assignatis suis vel heredibus eorum contra omnes homines warantizabimus
sicut donatores nostri predictam terram nobis
warantizabunt Ut autem hec nostra donacio et concessio rata et stabilis
imposterum permaneat hanc presentem cartam sigillo nostro roboravimus
Hiis testibus Willelmo Brian Astino filio Alicie
prepositis Burgi Anglico de Notinga' anno regni Regis Henrici filii
Johannis Regis !xv! Henrico Kytte Henrico le Taylur Augustino clerico et
aliis}.

\section{Previous work}
\label{SectionPrevious}

In this section we describe some previous work
on the problem of calendaring undated English charters
that had been carried out using the DEEDS data set.
First, however, we define some basic terms and set out the notation
that we will adhere to throughout.

We will use ${\cal D}$ to denote a generic text document; ${\cal D}$
will frequently be considered to be random---a selection from an
effectively infinite collection of documents that could have arisen in
the relevant random experiment.
Our data corpus will typically be denoted as
${\cal D}_1,{\cal D}_2, \ldots, {\cal D}_n$;
our notation will not distinguish whether these represent
random documents or their actual realizations,
as this will always be clear from the context.

A document ${\cal D}$ consists of a string (ordered sequence)
of not necessarily distinct words $(w_1, w_2, \ldots, w_m)$,
where $N({\cal D}) \equiv|{\cal D}| = m$
is the length of the document.
A \textit{shingle} of size $k$, or $k$-shingle, is a substring
$s_k = (w_{j+1}, w_{j+2}, \ldots, w_{j+k})$
of $k$ consecutive words in ${\cal D}$;
here $0 \leq j \leq m-k$ so there are $m-k+1$
(not necessarily distinct) $k$-shingles in ${\cal D}$.
We will let $s_k(\cal D)$ denote the set of these
(not necessarily distinct) $k$-shingles,
while $S_k(\cal D)$ will denote the set of distinct
$k$-shingles of ${\cal D}$.
The cardinalities of these sets is
$|S_k({\cal D})| \leq|s_k({\cal D})| = m-k+1$.
When $k$ is considered to be fixed,
and given a $k$-shingle $s \in s_k({\cal D})$,
we will let $n_s({\cal D})$ denote the number of times
this shingle occurs in $s_k({\cal D})$;
Finally, the date, $t$, of a document will be
denoted by $t({\cal D}) = t$.

Turning now to previous work on the DEEDS data, Rodolfo Fiallos worked
for the DEEDS project for many years, during which time he devised a
method for dating the manuscripts called the MT method. See \citet{Fia00}. MT stands for \textit{Multiplicador Total} in Spanish and
translates into English as ``Total Multiplier.'' Fiallos' method is
based on \textit{matching patterns}---shingles of arbitrary
length---which occur in the document we seek to date and which occur
also in one or more of the documents in a training set of dated
documents. The underlying idea is that a relatively higher
concentration of matching patterns should be found among those
documents in the training set whose dates are closer to the unkown date
of the document whose date we are trying to estimate. Fiallos
identified three characteristics of matching patterns thought to be
important for the calendaring process:

\textit{Length}: The number of words in the matching pattern
(shingle length).

\textit{Lifetime}: The difference, in years, between the last
and first occurrence of the matching pattern in the training set.
(If a pattern occurs only within one year, its Lifetime${}={}$0.)

\textit{Currency}: The Lifetime of the matching pattern
divided by the number of distinct years in which
it occurs.
(Here we are following the definition of R. Fiallos:
thus higher values of currency correspond to
sparser occurrence of the pattern throughout the years of its lifetime.)

To date a given document $\cal D$,
every substring of consecutive words
in $\cal D$ is examined.
[If $\cal D$ has length $m$, there will be
$m + (m-1) + \cdots+ 2 + 1 = m(m+1)/2$
such substrings in all.]
If such a substring occurs also in the training set
(it becomes a ``matching pattern'' and)
it produces an MT value defined as
\[
\mathrm{MT} = M_1(\mathrm{Length}) \times M_2(\mathrm{Lifetime})
\times M_3(\mathrm{Currency}).
\]
The larger its MT value, the more influential the
matching pattern is considered to be for the calendaring process.
Here the function $M_1$ is increasing since longer patterns
are considered to be more informative;
$M_2$ is decreasing since patterns with longer lifetimes
are viewed as being less informative;
and finally $M_3$ is also decreasing since sparser occurrence
of a pattern within its lifetime is
thought to reduce its evidentiary worth.
The functions $M_1$, $M_2$ and $M_3$
can be defined in many \textit{ad hoc} ways,
and such definitions invariably entail many tuning-type parameters;
such functions and their parameters were determined by Fiallos
through extensive trial and error
and leave-one-out cross-validation.

Once MT values have been assigned to all
matching patterns in $\cal D$,
an MT value is computed for every year
for which training data is available
by summing the MT values of all of the matching patterns
of $\cal D$ that occur among the training data of that year.
However, in an attempt to reduce noise,
matching patterns whose MT values fall below
a certain threshold are excluded
from this summing process.
This procedure leads to a function of time,
called the MT function.
To account for the fact that the number
of training documents varies over time,
the values of this MT function are each divided
by the number of training documents in that year.
These standardized values are referred to as Global MT or GMT values.
In principle, the date having the highest GMT value
is taken to be the estimated date of $\cal D$.
However, because such GMT functions are still quite noisy,
the GMT values are first averaged over time intervals of,
say, 40 or 20 years, leading to an estimated time interval
for the date of $\cal D$.
This estimated date range is then expanded,
and the GMT averaging process is then repeated over this new range
but now using a smaller interval width.
This process is repeated several times,
leading finally to a point estimate for the unknown date.

\begin{figure}

\includegraphics{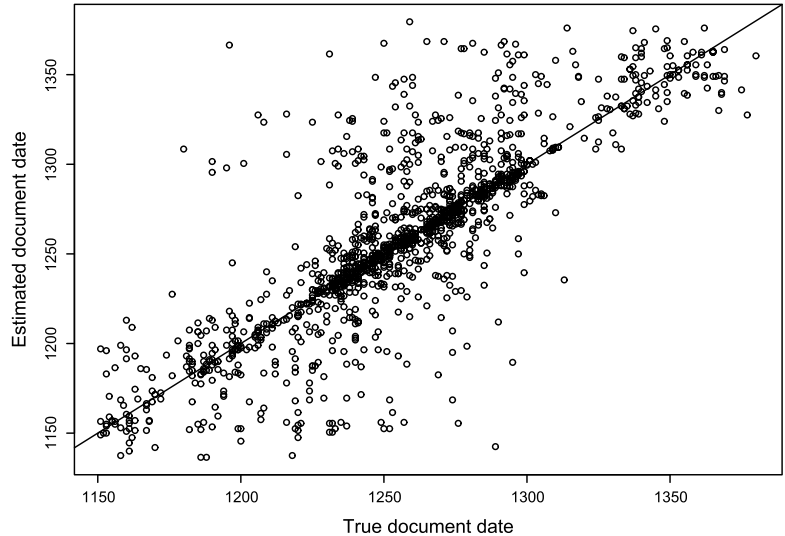}

\caption{Estimated versus true dates
for 1484 documents, dated by the method of R. Fiallos,
each selected randomly from a training set
of approximately 3500 dated documents.}\label{figur3}
\end{figure}

Figure \ref{figur3} (based on computations provided by Fiallos) plots
the estimated versus the actual dates for 1484 DEEDS documents which
were dated by the MT method. These 1484 documents were randomly
selected from a set of approximately 3500 dated documents, and each of
these 1484 selected documents was then dated on the basis of the full
3500 documents data set, but with the one being dated left out. The
mean absolute error (MAE) was found to be 16 years. The heavy
concentration of points occurring near the ``$x=y$'' axis is due to
documents that have been dated rather accurately. We remark, however,
that the MAE estimate of 16 years is likely to be optimistic because it
was not based on a held-out test set---that is, the optimization of the
many tuning parameters was performed over the same data set.

\section{Calendaring by nearest neighbors (kNN)}
\label{SectionNearest}

Distance based methods for calendaring charters (also referred to as
nearest neighbor or kNN methods) were introduced in Feuerverger et~al.
(\citeyear{Feuetal05}, \citeyear{Feuetal08}), hereafter referred to as
FHTG (\citeyear{Feuetal05}) and  FHTG (\citeyear{Feuetal08}). The
underlying idea is to define measures of \textit{distance} between
pairs of documents and to estimate the date of an undated document by a
weighted average of the dates of documents in a training set using
weights which depend on their distances to the document we seek to
date. Alternately, one can use a reciprocal to the concept of distance,
namely, \textit{similarity} (also referred to as \textit{resemblance}
or \textit{correspondence}), and average over the dates of documents in
the training set using weights based on the similarity measures. For
completeness and later comparisons, we outline these methods in this
section.

\textit{Measures of distance and similarity}:
Distance and similarity measures on documents are discussed,
for example, in \citet{Dje03}, FHTG (\citeyear{Feuetal05}),
\citet{McGKolNor}, \citet{Quaetal},
\citet{SalWanYan75}, \citet{TanSteKum05}, \citet{ZhaKor99}
and references therein.
Let $\cal P$ and $\cal Q$ represent two documents
whose union consists
of $| {\cal P} \cup{\cal Q}| = \ell$ distinct words.
(A discussion based on $k$-shingles would be analogous.)
Let $p \equiv(p_1, \ldots, p_\ell)$
and $q \equiv(q_1, \ldots, q_\ell)$,
respectively,
be vectors corresponding to the occurrence of these distinct words;
these vectors can variously be word counts,
normalized counts ($\sum_i p_i = \sum_i q_i= 1$)
or 0--1 incidence vectors.
Then some natural measures of \textit{similarity}
between $\cal P$ and $\cal Q$ are given by
%
\begin{equation}\label{typeonesimilarity}
\mathrm{Sim}_{\gamma} ({\cal P}, {\cal Q}) = \frac{\sum_{i = 1}^\ell
p_i^\gamma q_i^\gamma} {
\sqrt{\sum_{i = 1}^\ell p_i^{2\gamma}} \sqrt{\sum_{i = 1}^\ell
q_i^{2\gamma}}}
\end{equation}
for $0 < \gamma< \infty$.
The case $\gamma= 1$ corresponds to the
angle-based \textit{cosine similarity},
while the case $\gamma= 1/2$ with normalized $p$ and $q$
results in a similarity measure that leads
to a \textit{Hellinger distance}.
Similarity measures somewhat alike to (\ref{typeonesimilarity})
may also be defined as
%
\begin{equation}\label{typetwosimilarity}
\mathrm{Sim}_{\alpha} ({\cal P}, {\cal Q}) = \frac
{ \sum_1^\ell p_i^\alpha q_i^\alpha} {
\sum_1^\ell( p_i^{2\alpha} + q_i^{2\alpha} - p_i^\alpha
q_i^\alpha) }
\end{equation}
for $0 < \alpha< \infty$.
Unlike (\ref{typeonesimilarity}), these have the advantage that,
for all such values of~$\alpha$,
%
\begin{equation}\label{typetwodistance}
\mathrm{Dist}_\alpha({\cal P}, {\cal Q}) \equiv1 -
\mathrm{Sim}_{\alpha} ({\cal P}, {\cal Q})
\end{equation}
is a proper metric (i.e., satisfies the triangle inequality).

\citet{Bro98} defined the \textit{resemblance}
of two documents ${\cal D}_1$ and ${\cal D}_2$,
for a given (fixed) shingle size $k$, as
%
\begin{equation}\label{Broder}
\mathrm{Res}_k({\cal D}_1, {\cal D}_2) \equiv
\frac
{| {\cal S}_k({\cal D}_1) \cap{\cal S}_k({\cal D}_2) |} {
| {\cal S}_k({\cal D}_1) \cup{\cal S}_k({\cal D}_2) |}.
\end{equation}
Using this definition, a set-based \textit{resemblance distance}
between documents which satisfies
the triangle inequality may be defined as
\[
\mathrm{Dist}_k({\cal D}_1, {\cal D}_2)
\equiv1 - \mathrm{Res}_k({\cal D}_1, {\cal
D}_2).
\]

There are, of course, may other measures of distance and similarity.
We remark that for information retrieval work,
many distance measures often behave similarly
and that whether or not the triangle inequality holds
tends to be inconsequential.
[See, e.g., \citet{Dje03}, Chapter 4.]
One potential benefit, however, of having many
versions of distance is in permitting the
implementation of ensemble-type estimation methods.
The use of multidimensional scaling as an alternative
to incorporate distances based on similarities is
also worth mentioning, but lies outside the scope
of this paper.

\textit{Calendaring by kNN methods}:
To develop and evaluate
distance based and other estimation methods,
the DEEDS documents were first partitioned at random
into a \textit{training} set ${\cal T}$,
a \textit{validation} set ${\cal V}$ and a \textit{test} set ${\cal A}$.
We will frequently interchange notation such as
${\cal D}_i \in{\cal T}$ and $i \in{\cal T}$
for membership in these sets.
Our aim is to estimate the unknown date $t_i$
of a document ${\cal D}_i$, when $i \notin{\cal T}$.
Here we follow FHTG (\citeyear{Feuetal05}, \citeyear{Feuetal08}).

Let $d_k(i,j)$, for $k = 1, 2, \ldots, r$,
denote $r$ different distance measures between documents
${\cal D}_i$ and ${\cal D}_j$, say.
For instance, these distances could all be Broder distances
corresponding to different shingle lengths $k$,
with $r$ being the largest shingle size in the procedure.
Using these distances, we define an $r$-dimensional
kernel weight on the dates $t_j$
of the documents ${\cal D}_j$ in the training set ${\cal T}$:
%
\begin{equation}\label{aij}
a(i,j) \equiv a ( i,j | h_1, \ldots, h_r ) = \prod
_{k=1}^r K_{h_k} \bigl(
d_k(i,j) \bigr),
\end{equation}
where $i$ corresponds to the document ${\cal D}_i$
we seek to date.
Here $K_h (\cdot)$ is a nonnegative, nonincreasing function
defined on the positive half-line
and $h$ is a bandwidth parameter.
For example, we could take
$K_h (u) \propto\exp\{ -(u/h)^2 \}$,
or $K_h (u) \propto( 1 + (u/h)^2 )^{-\eta}$
for some choice of $\eta$,
with each distance measure permitted to have its own bandwidth.
The distance based (or kNN) estimator for the date $t_i$
of ${\cal D}_i$ is then defined as
%
\begin{equation}\label{Hallestimator}
\hat t \equiv\hat t_i \equiv\mathop{\arg\min}_t \sum
_{j \in{\cal T}} ( t_j - t )^2 a(i,j) =
\frac{\sum_{j \in{\cal T}} t_j a(i,j)} {
\sum_{j \in{\cal T}} a(i,j)}.
\end{equation}

It remains to consider the selection of the bandwidths
$h_1, \ldots, h_r$ in (\ref{aij}).
In FHTG (\citeyear{Feuetal05}, \citeyear{Feuetal08}) this
was based on a form of cross-validation which is local
in the sense that it tries to determine
the set of bandwidths optimal
for each document ${\cal D}_i$ individually.
Specifically, let ${\cal K}(i)$ be the collection
of nearest neighbors to ${\cal D}_i$,
defined as the union, over all $1 \leq k \leq r$,
of the set of all indices $j \in{\cal T}$ in the training set
such that $d_k (i,j)$ is among the $m$ smallest values
of that quantity,
where the integer $m$ is some small fraction
of the total number of documents in~$\cal T$.
Then $m$, as well as the $h_1, \ldots, h_r$
specific to ${\cal D}_i$,
are chosen to minimize the cross-validation function
%
\begin{equation}\label{TheCVcriterion}
\operatorname{CV} (m; h_1, \ldots, h_r) = \frac{1}{{ | \cal K}(i) |} \sum
_{j^\prime\in{\cal K}(i)} ( t_{j^\prime} - \hat t_{ - j^\prime}
)^2,
\end{equation}
where
\begin{eqnarray*}
\hat t_{ - j^\prime} &=& \hat t_{ - j^\prime} (m; h_1, \ldots,
h_r) = \mathop{\arg\min}_{t} \sum_{j \in{\cal T}, j \neq j^\prime}
( t_j - t )^2 a\bigl( j^\prime, j\bigr)
\\
&=& \frac
{\sum_{j \in{\cal T}, j \neq j^\prime} t_j a(j^\prime, j)} {
\sum_{j \in{\cal T}, j \neq j^\prime} a(j^\prime, j)}.
\end{eqnarray*}
While this bandwidth selection process is local
in the sense that for each ${\cal D}_i$,
it tries to determine a set of bandwidths
by optimizing over its nearest neighbors~${\cal K}(i)$,
if we were to choose all ${\cal K}(i) \equiv{\cal T}$
the procedure would become global
with the estimated bandwidths then being the same
for all of the documents.

The optimization over $m$ and $h_1, \ldots, h_r$
is carried out via a grid search resulting in
\[
(\hat m; \hat h_1, \ldots, \hat h_r) = \arg\min
\operatorname{CV} (m; h_1, \ldots, h_r).
\]
The mean squared error of the date estimate $\hat t_i$
can then be estimated as
\[
\hat s^2(i) = \frac
{\sum_{j^\prime\in{\cal K}(i)}
( t_{j^\prime} - \hat t_{- j^\prime} )^2
a ( i,j^\prime | \hat h_1, \ldots, \hat h_r )} {
\sum_{j^\prime\in{\cal K}(i)}
a ( i,j^\prime | \hat h_1, \ldots, \hat h_r ) },
\]
where the $\hat t_{-j'}$, for all $j' \in{\cal K}(i)$,
are computed using the same bandwidths as for~$\hat t_i$.

\section{Calendaring by maximum prevalence (MP)}
\label{SectionMaximum}

Our method of maximum prevalence (MP)
for calendaring a document $\cal D$
is an analogue of the method of maximum likelihood;
it attempts to assign, for each point $t$ in time,
a probability for the occurrence of $\cal D$ at that time,
and it estimates the unknown date of $\cal D$ by that value of $t$
at which $\cal D$ has the highest probability of occurrence.
The MP method is specific to a given shingle size, say, $k$,
but the ensemble of estimates produced
using different values of $k$
can subsequently be combined.

If now $\cal D$ consists of a string of $N$ words, it will contain
$\llvert s_k ({\cal D}) \rrvert = N -k +1$ (not necessarily unique)
$k$-shingles. We will let $N({\cal D}) \equiv\llvert s_k ({\cal D})
\rrvert$ represent the number of elements in $s_k ({\cal D})$,
suppressing its dependence on $k$. The assumption is then made that
these $N({\cal D})$ shingles occur independently of each other and are
drawn from the multivariate distribution over shingles of size $k$ in
effect at the true date $t({\cal D})$ of the document. Although this
assumption---made here of necessity---is untrue, there are some
arguments in its favor. In particular, in some statistical problems,
estimators can remain consistent (and even asymptotically efficient)
when dependency is ignored. Examples include incorrectly assuming
independence when estimating the mean of certain stationary processes.
In such cases, it is primarily the variances of the estimates that are
affected. Additional arguments are given in \citet{DomPaz96}.

Suppose then that for every possible $k$-shingle $s$,
we knew the probability $\pi_s(t)$ of its occurrence at every time
point $t$.
Then the \textit{prevalence function} for $\cal D$ is defined as
%
\begin{equation}\label{prevalence}
\pi_{{\cal D}}(t) = \prod_{ s \in s_k ({\cal D})}
\pi_s(t),
\end{equation}
and by analogy with maximum likelihood,
the true date $t({\cal D})$ of $\cal D$
would be estimated as that value of $t$
at which $\pi_{{\cal D}}(t)$ is maximized.
The function $\pi_{{\cal D}}(t)$
is intended to represent the probability
of the occurrence of $\cal D$ as a function over time.
Of course, we do not know the $\pi_s(t)$,
but these may be estimated, as $\hat\pi_s(t)$, say,
leading to an estimated prevalence function
%
\begin{equation}\label{estimatedprevalence}
\hat\pi_{{\cal D}}(t) = \prod_{ s \in s_k ({\cal D})} \hat
\pi_s(t),
\end{equation}
and finally to our proposed date estimator
\[
\hat t_{{\cal D}} = \mathop{\arg\max}_t \hat\pi_{{\cal D}}(t).
\]

We must now consider how to estimate the probabilities
$\pi_s(t)$ of shingle occurrence.
Given a document ${\cal D}$ and a $k$-shingle $s$,
the number of times $s$ occurs in ${\cal D}$
will be denoted by $n_s({\cal D})$.
For $n_s({\cal D})$ we postulate the binomial model
\[
{\cal L} \bigl( n_s({\cal D}) | N({\cal D})=N, t({\cal D})=t \bigr)
\sim \operatorname{Bin} \bigl( N, \pi_s(t) \bigr)
\]
according to which the probability of the observed value $n_s({\cal
D})$ is
\[
\pmatrix{ {N({\cal D})}
\cr
{n_s({\cal D})} } \bigl\{
\pi_s(t) \bigr\}^{n_s({\cal D})} \bigl\{ 1 - \pi_s(t)
\bigr\}^{N({\cal D}) - n_s({\cal D})};
\]
here $t({\cal D}) = t$ is the date of ${\cal D}$
and $N({\cal D}) = N$ is the number of $k$-shingles it contains.
In terms of the canonical log-odds parameter
\[
\lambda_s(t) \equiv\log\frac{\pi_s(t)}{1 - \pi_s(t)},
\]
the logarithm of this probability is
\[
\log\pmatrix{ { N({\cal D})}
\cr
{n_s({\cal D}) } } +
n_s({\cal D}) \lambda_s(t) - N({\cal D}) \log\bigl[ 1
+ \exp\bigl\{ \lambda_s(t) \bigr\} \bigr].
\]
Because the first (combinatorial) term here
does not depend on $\lambda_s(t)$,
we drop it from subsequent expressions.
Hence, given a random sample of documents \mbox{${\cal D}_i \in{\cal T}$},
with corresponding dates $t_i$, the log-likelihood function
in the parameter $\lambda_s(\cdot)$ is taken to be
\[
\sum_{i \in{\cal T}} \bigl\{ n_s({\cal
D}_i) \lambda_s(t_i) - N({\cal
D}_i) \log\bigl[ 1 + \exp\bigl\{ \lambda_s(t_i)
\bigr\} \bigr] \bigr\}.
\]
We next model the function parameter $\lambda_s(\cdot)$
as a $t$-local polynomial of degree~$p$;
specifically, for $u$ near $t$,
%
\begin{equation}\label{polynomial}
\lambda_s(u) \approx\beta_0 + \beta_1 (u
- t) + \cdots+ \beta_p (u-t)^p.
\end{equation}
Here the dependence of $\lambda_s (\cdot)$ as well as of the
$\beta_0, \ldots, \beta_p$ on $t$ has been suppressed.
[See, e.g., \citet{Loa99}.]
Finally, we\vadjust{\goodbreak} introduce a $t$-localized version
of the log-likelihood, namely,
%
\begin{equation}\label{locallikelihood}
\sum_{i \in{\cal T}} \bigl\{ n_s({\cal
D}_i) \lambda_s(t_i) - N({\cal
D}_i) \log\bigl[ 1 + \exp\bigl\{ \lambda_s(t_i)
\bigr\} \bigr] \bigr\} K_{h} (t_i - t),
\end{equation}
which is to be maximized over the $\beta_0, \ldots, \beta_p$
for every given $t$.
The resulting estimate $\hat\beta_0$ for $\beta_0$ is then taken
as our estimate for $\lambda_s(t)$.
Here $K_{h}(u)$ is a symmetric weight function
which takes on its maximum at $u=0$,
and is nonincreasing as $u$ moves away from the origin.
A Gaussian version might be
$K_{h}(u) \propto\exp\{ -u^2 / 2 h^2 \}$,
with $h$ corresponding to its standard deviation.
More flexibly, we could write $K_{h,\nu}(u)$
in place of $K_{h}(u)$, with
\[
K_{h,\nu}(u) \propto\biggl( 1 + \frac{u^2}{\nu h^2} \biggr)^{-(\nu+1)/2}
\]
corresponding to a $t$-distribution with $\nu$ degrees of freedom;
this allows for a tail-weight parameter in addition to a scaling.

If we take the polynomial (\ref{polynomial}) to be of degree $p=0$,
so that $\lambda_s (u) = \beta_0$ there,
and then set the derivative with respect to $\beta_0$
in (\ref{locallikelihood}) to zero,
we obtain (in terms of $\pi_s$) the solution
%
\begin{equation}\label{NadarayaWatson}
\hat\pi_s(t) = \frac
{ \sum_{i \in{\cal T}} n_s({\cal D}_i) K_h (t_i - t) } {
\sum_{i \in{\cal T}} N({\cal D}_i) K_h (t_i - t) },
\end{equation}
which is analogous to the estimator of \citet{Nad64} and \citet{Wat64}.
If instead we use a polynomial of degree $p=1$ in (\ref{polynomial})
(locally linear smoothing)
and set derivatives with respect to $\beta_0$ and $\beta_1$
in (\ref{locallikelihood}) to zero,
we obtain the pair of equations
%
\begin{equation}\label{LocalLinearOne}\qquad
\sum_{i \in{\cal T}} n_s({\cal D}_i)
K_h (t_{{\cal D}_i} - t) = \sum_{i \in{\cal T}}
\frac
{ N({\cal D}_i)
\exp\{ \beta_0 + \beta_1 (t_{{\cal D}_i} - t) \} } {
1 + \exp\{ \beta_0 + \beta_1 (t_{{\cal D}_i} - t) \} } K_h (t_{{\cal
D}_i} - t)
\end{equation}
and
%
\begin{eqnarray}\label{LocalLinearTwo}
&&
\sum_{i \in{\cal T}} n_s({\cal D}_i)
(t_i - t) K_h (t_i - t)\nonumber\\[-8pt]\\[-8pt]
&&\qquad = \sum
_{i \in{\cal T}} \frac
{ N({\cal D}_i)
\exp\{ \beta_0 + \beta_1 (t_i - t) \} } {
1 + \exp\{ \beta_0 + \beta_1 (t_i - t) \} } (t_i - t)
K_h (t_i - t).\nonumber
\end{eqnarray}
These must be solved numerically for $\beta_0$ and $\beta_1$
at every $t$, giving $\hat\beta_0$ and $\hat\beta_1$,
and we would then take
\[
\hat\pi_s(t) = \frac{\exp(\hat\beta_0) }{ 1 + \exp(\hat\beta_0)}.
\]

We remark that we could alternatively have modeled the data
using a poisson distribution as in
\[
n_s({\cal D}) \sim \operatorname{Poisson} \bigl( \mu_s(t) N({
\cal D}) \bigr)\vadjust{\goodbreak}
\]
and carried out local polynomial fitting using the canonical
link parameter $\lambda_s(t) = \log\mu_s(t)$.
[Here we have used $\mu_s(t)$ in place of $\pi_s(t)$
for the shingle's probabilities.]
If the local polynomial is taken to be of degree 0,
this leads again to the
Nadaraya--Watson type solution (\ref{NadarayaWatson}),
with $\hat\mu_s(t) = \hat\pi_s(t)$.
For local polynomials of degree greater than 0,
the solutions are approximately,
but not exactly, equivalent to the binomial case.
Note that due to their exponential family nature,
the Hessians associated with these models
are strictly negative definite; hence, these various equations
are well-behaved and have unique solutions.

As a final remark, we mention that one may consider
replacing the definition of the prevalence function
in (\ref{prevalence}) by something like
%
\begin{equation}\label{prevalencetwo}
\pi_{{\cal D}}(t) = \prod_{ s \in s_k ({\cal D})}
\pi_s(t) \prod_{ s \notin s_k ({\cal D})} \bigl[ 1 -
\pi_s(t) \bigr]
\end{equation}
with a corresponding change in its empirical version
(\ref{estimatedprevalence}),
so as to try to take into better account shingles that
did \textit{not} occur in the document being dated.
However, the logarithm of the second factor
in (\ref{prevalencetwo}) is
%
\begin{equation}
\sum_{ s \notin s_k ({\cal D})} \log\bigl\{ 1 - \pi_s(t)
\bigr\} \approx- \sum_{ s \notin s_k ({\cal D})} \pi_s(t)
\approx- \sum_{ s} \pi_s(t) = -1,
\end{equation}
since each $\pi_s(t)$ is small,
and because the total number of possible shingles
far exceeds those in any given document.
We computed empirical versions of the logarithm
of the second factor in (\ref{prevalencetwo})
and invariably found that such curves
stayed close to $-1$, and were therefore not informative.

\section{Calendaring via quantile regression (QR)}
\label{SectionThird}

A third proposal for the calendaring problem
is based on quantile regression as follows.
Suppose that $\cal D$ is a document
whose date we wish to estimate.
A scatterplot is produced of the distances
$\operatorname{Dist} ( {\cal D}, {\cal D}_i )$
from $\cal D$ to each of the documents
${\cal D}_i \in{\cal T}$ in a training set,
against the known dates $t ( {\cal D}_i )$
of those training set documents.
A nonparametric quantile regression (QR) curve
is then fit to this scatterplot,
and the date at which this QR plot attains its minimum value
is taken as the estimate of the date of $\cal D$.
QR algorithms typically have two parameters:
a bandwidth $h$ which controls the smoothness of the curve
and a quantile $0 < q < 1$.
(The bandwidth parameter need not be kept constant
over the range of dates
and may be larger in regions of sparser date ranges.)
The parameters $h$ and $q$ are meant to be optimized
for documents in a validation set which are dated
using data in a training set.
The procedure is then assessed on the documents
in a held-out test set.
Figure \ref{FigQR} in Section~\ref{SectionResults} below
illustrates the QR procedure in action.
For quantile regression, our key reference is \citet{Koe05}.

\section{Theoretical considerations}
\label{SectionTheory}

In this section we discuss some general considerations
concerning the consistency of the estimates proposed in
Sections \ref{SectionNearest} and~\ref{SectionMaximum}.

Turning first to the distance-based (kNN) method,
we have the following result:
Let ${\cal D}_0$ be an undated document written at time $t_0$,
and denote by ${\cal D}$ a dated document,
written at time $T$,
and chosen at random from a potentially infinite
(but representative) training set
and having a random distance $\Delta$ from ${\cal D}_0$.
(For simplicity, we assume
that our kNN procedure is based on only
a single distance measure, but the general case is similar.)
We posit five conditions:

\begin{longlist}
\item
\textit{Asymptotic unbiasedness}:
The conditional expectation of the mean of $T$ converges to $t_0$
over neighborhoods $\Delta\rightarrow0$.

\item \textit{Bounded variance}:
The second moment of $T$ remains bounded
as these distance neighborhoods shrink to 0.

\item \textit{A technical condition}:
$\Delta$ can be viewed as possessing a density at the origin
which is continuous and positive.

\item The kernel $K(u)$ is bounded, continuous,
compactly supported and nonincreasing on the positive
real line, with $K(0) > 0$.

\item The number of elements in the training set
increases sufficiently quickly as the bandwidth
$h$ tends to 0.
\end{longlist}
Under the conditions (i)--(v), it was proved in FHTG (\citeyear{Feuetal08})
that the estimator $\hat t$ defined at (\ref{Hallestimator})
is a consistent estimator of the true date $t_0$
of the document ${\cal D}_0$, that is, $\hat t \rightarrow_p t_0$
as the size of the training set tends to infinity.

Turning to the MP method,
consistency results may be established
along the following lines.
Assume time to be integer valued
and restricted to a compact domain:
$t_{\mathrm{min}} \leq t \leq t_{\mathrm{max}}$.
We again let ${\cal D}_0$ denote the document to be dated,
and $t_0$ is its unknown true date.
We consider our training set to be an
increasing ($n \rightarrow\infty$) sequence of documents
${\cal T}_n \equiv\{ {\cal D}_1, {\cal D}_3, \ldots, {\cal
D}_n \}$
in which the random documents~${\cal D}_i$,
and their corresponding random dates $T_i$,
are viewed as being an i.i.d. sample from an
essentially infinite population generically represented
by the random object $({\cal D}, T)$.
The set of all shingles possible at any point within our time interval
will be denoted by ${\cal S}$.
(The shingle size is considered fixed.)
Every shingle $s \in{\cal S}$ has associated with it
its probabilities $\pi_s(t)$ of being drawn
at any of the time points $t$.
Note that for each $t$ we will have
$\sum_{s \in{\cal S}} \pi_s(t) = 1$.
We next assume that if $t_1 \neq t_2$, then the collections
$ \{ \pi_s(t_1); s \in{\cal S} \}$ and
$ \{ \pi_s(t_2); s \in{\cal S} \}$ are not identical;
specifically, $\pi_s(t_1) \neq\pi_s(t_2)$ for some $s \in{\cal S}$.

In the random object $({\cal D}, T)$,
we will assume that, conditionally on $T = t$,
the sequence of shingles comprising ${\cal D}$
is an i.i.d. sample drawn from ${\cal S}$
under the probability distribution
$ \{ \pi_s(t)\dvtx s \in{\cal S} \}$.
In particular,
the shingles of ${\cal D}_0$ are assumed
to be randomly drawn from ${\cal S}$
using the distribution $\pi_s(t_0)$.
In $({\cal D}, T)$ the length of ${\cal D}$
is assumed to be independent of $T$.

Now, for each $s \in{\cal S}$,
under standard conditions for the Nadaraya--Watson estimator,
we will have
%
\begin{equation}
\sup_{t_{\mathrm{min}} \leq t \leq t_{\mathrm{max}}} \bigl\llvert\hat\pi_s(t) -
\pi_s(t) \bigr\rrvert\rightarrow0\qquad \mbox{as } n \rightarrow\infty,
\end{equation}
so that for a ${\cal D}_0$ of fixed, finite length, we will have
%
\begin{equation}
\sup_{t_{\mathrm{min}} \leq t \leq t_{\mathrm{max}}} \bigl\llvert\hat\pi_{{\cal
D}_0}(t) -
\pi_{{\cal D}_0}(t) \bigr\rrvert\rightarrow0\qquad \mbox{as } n \rightarrow
\infty.
\end{equation}
On the other hand, the standard argument
(based on the Law of Large Numbers and Kullback--Leibler distance)
which is used to prove consistency of the MLE
in the case when the parameter space is finite
applies equally here and allows us to conclude
that with arbitrarily high probability, $\pi_{{\cal D}_0}(t)$
will take on its maximum value uniquely at $t_0$ provided only that
$ \llvert {\cal D}_0 \rrvert$ is sufficiently large.
Hence, by requiring $ \llvert {\cal D}_0 \rrvert$ to be sufficiently large,
and then letting $n \rightarrow\infty$,
the estimated date $\hat t$ of ${\cal D}_0$ can be made to equal $t_0$
with arbitrarily high probability.

Of course, asymptotic results do need to be assessed
for relevance in any specific application.
In particular, it must be borne in mind
that any document to be dated will be of finite length
and so will necessarily contain only limited ``Fisher information''
for the estimation of its date parameter.

\section{Numerical work}
\label{SectionResults}

In this section we describe some numerical experiments
which we conducted using the kNN and MP estimation methods
with the DEEDS data set.
This work was carried out using a combination
of UNIX commands together with the C programming language,
as well as the R statistical computing package.

For the purposes of our experiments,
we first randomly partitioned the 3353 DEEDS documents
which were available to us into a \textit{training} set ${\cal T}$,
a \textit{validation} set ${\cal V}$
and a \textit{test} set ${\cal A}$,
with these sets having cardinalities $|{\cal T}| = 2608$,
$|{\cal V}| = 419$, and $|{\cal A}| = 326$.
Unlike the MP method, however,
our experiments with the kNN method
as described in Section \ref{SectionNearest}
did not require a validation set
because in that method the parameters for dating any given document
are determined solely from its neighbors
within the training set, as well as from
other members of the training set.
Therefore, for our kNN numerical work, ${\cal V}$ and ${\cal A}$
were combined to form a larger test set consisting of 745 documents.

Our experiments with the kNN method
were based on shingle sizes 1, 2 and 3,
as well as on all combinations of these sizes.
We used the distance (\ref{typetwodistance})
with $\alpha= 1$,
based on the similarity (\ref{typetwosimilarity}),
and therefore a proper metric;
this distance was computed using argument vectors
``$p$'' and ``$q$'' consisting of raw (i.e., unnormalized) shingle counts.
These distances are denoted as $d_k(i,j)$,
with $k$ representing the shingle sizes on which they are based.

For a given document ${\cal D}_i$ in our test set of 745 documents,
all 2608 of its distances to the documents in the training set were computed
for each of the three shingle sizes.
The set ${\cal K}(i)$ of neighbors to ${\cal D}_i$
was formed by taking all indices $j \in{\cal T}$
such that $d_k(i,j)$ is among the $m$ smallest values
of that distance.
When multiple shingle sizes were used,
the set of neighbors ${\cal K}(i)$ was taken to be
the union over the $m$ smallest distances
for each of the shingle sizes used.
As the kNN procedure was not very sensitive
to the exact choice of $m$,
the values of $m$ we experimented with were 5, 10, 20, 100, 500 and 1000.
(The smaller $m$ values, of course, result in faster computation times.)
The optimal bandwidths for use with ${\cal D}_i$
were then determined (entirely from within the training set)
using the procedure defined at (\ref{TheCVcriterion})
together with a standard Gaussian kernel at (\ref{aij}).
For each $m$ and~${\cal D}_i$, these bandwidths were determined
by searching over a one-, two- or three- dimensional grid,
depending on the number of shingle lengths used in the procedure;
the optimal bandwidths so resulting were
therefore different for each~${\cal D}_i$.
Finally, we computed the RMSE (root mean squared error),
MAE (mean absolute error)
and MedAE (median absolute error) performance measures
for the resulting date estimates of the 745 documents
in our (enlarged) test set.

\begin{table}
\caption{Performance of the kNN and MP methods
on the DEEDS data set}\label{GelilaA}
\begin{tabular*}{\tablewidth}{@{\extracolsep{\fill}}lclccc@{}}
\hline
\textbf{Dating}& \multicolumn{1}{c}{\textbf{Shingle}}
& \multicolumn{1}{c}{\textbf{Optimal}} &
\multicolumn{1}{c}{$\bolds{\sqrt{\mathrm{MSE}}}$}
& \multicolumn{1}{c}{\textbf{MAE}} &
\multicolumn{1}{c@{}}{\textbf{MedAE}} \\
\textbf{method} & \multicolumn{1}{c}{\textbf{lengths}}
& \multicolumn{1}{c}{\textbf{parameters}} &
\multicolumn{1}{c}{\textbf{(val., test)}}
& \multicolumn{1}{c}{\textbf{(val., test)}}
& \multicolumn{1}{c@{}}{\textbf{(val., test)}} \\
\hline
M1 & 1 & $h=8$, $\mbox{df}=5$ & 18.3, 19.8 & 11.7, 12.5 & 7.0, 8.0 \\
M2 & 2 & $h=12$, $\mbox{df}=3$ & 14.8, 14.7 & \hphantom{0}9.5, 9.0\hphantom{0} & 6.0, 6.0 \\
M3 & 3 & $h=12$, $\mbox{df}=5$ & 17.0, 15.4 & 10.1, 9.5\hphantom{0} & 6.0, 6.0 \\
M4 & 4 & $h=16$, $\mbox{df}=12$ & 18.8, 22.8 & 11.5, 12.4 & 7.0, 7.0 \\
M1234& 1--4 & \multicolumn{1}{c}{---} & 14.3, 14.5 & \hphantom{0}9.3, 9.2\hphantom{0} & 6.0, 6.0 \\
[4pt]
kNN1 & 1 & \multicolumn{1}{c}{$m=1000$} & 20.1 & 12.3 & 6.4 \\
kNN2 & 2 & \multicolumn{1}{c}{$m=500$\hphantom{0}} & 23.7 & 13.8 & 6.4 \\
kNN3 & 3 & \multicolumn{1}{c}{$m=500$\hphantom{0}} & 28.3 & 16.6 & 7.6 \\
kNN12 & 1 \& 2 & \multicolumn{1}{c}{$m=100$\hphantom{0}} & 20.2 & 12.1 & 6.3 \\
kNN13 & 1 \& 3 & \multicolumn{1}{c}{$m=100$\hphantom{0}} & 21.7 & 12.9 & 7.0 \\
kNN23 & 2 \& 3 & \multicolumn{1}{c}{$m=100$\hphantom{0}} & 25.5 & 14.9 & 6.8 \\
kNN123 & 1 \& 2 \& 3 & \multicolumn{1}{c}{$m=10$\hphantom{00}} & 25.4 & 15.0 & 7.9 \\
\hline
\end{tabular*}
\end{table}

The results of these computations are summarized
in the last seven rows of Table~\ref{GelilaA},
labeled kNN1 (based on shingle size 1)
to kNN123 (based on using shingle sizes 1, 2 and 3 simultaneously).
Among these, the combination kNN12 is seen to be best,
although kNN1 performed similarly in terms of MAE,
while kNN1 and kNN2 performed similarly in terms of MedAE.
The optimal choices for $m$
are also shown in the table.
The apparent deterioration of performance for kNN123
appears related to the fact that
$m$ was held equal for all three shingle sizes.
The relatively large values of RMSE occur because
a small number of documents could not be dated at all accurately.
By way of comparison,
the mean year for the training documents was approximately 1246,
and if this value were used to estimate the dates
of the documents in the test set, the RMSE would be 47,
the MAE would be~37, and the MedAE would be 25.

\begin{figure}

\includegraphics{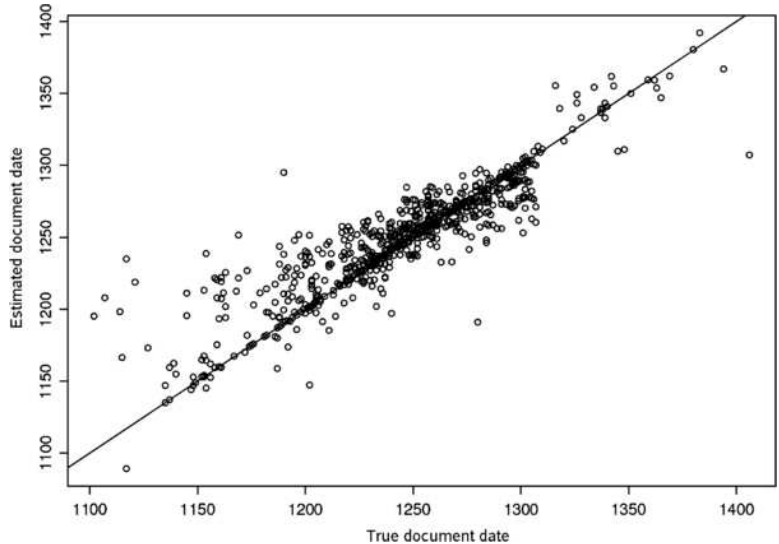}

\caption{Estimated versus true dates
for the 745 documents in the test set,
using the kNN method with $m = 100$
and combining shingle lengths 1 and 2.
The solid line is ``$y=x$.''}\label{FigurekNN}\vspace*{-3pt}
\end{figure}

Figure \ref{FigurekNN}, as an example, shows the estimated versus
the (presumed) true dates for the 745 documents
in the test set for the kNN12 procedure based on $m = 100$.
This figure evidences some degree of edge bias,
with early documents having overestimated dates
and later ones having somewhat underestimated dates.
This bias is due to the one-sided nature
of nearest neighbors at the edges.

Our experiments with the maximum prevalence (MP) methods
required all three of the sets ${\cal T}$, ${\cal V}$ and ${\cal A}$.
To save computational labor, we implemented only
the locally constant (i.e., Nadaraya--Watson type) version
(\ref{NadarayaWatson}) for estimating the
shingle probability functions;
we used the $t$-distribution kernel\vspace*{1pt}
$K(x) = ( 1 + x^2 / \nu)^{-(\nu+ 1)/2}$.
For each of the shingle sizes 1, 2, 3 and~4,
optimal values of the bandwidth $h$
and degrees of freedom parameter $\nu$
were determined by optimizing the date estimates for the documents
in the validation set using the training data.
Finally, the performance measures were computed
on both the validation and the test set
using the parameters that were determined on the validation set.
These results are shown on the rows labeled
M1, M2, M3 and M4 of Table \ref{GelilaA}.
For each of these methods, the optimized parameter values are shown,\vadjust{\goodbreak}
and the RMSE, MAE and MedAE performance measures are given
for both the validation and the test set data.
The best performing of these methods was that based
on shingle size~2 (i.e., method M2),
with a median absolute error of 6.0.
The shingle size 2 is, in some sense, the best compromise
(for a data set of this size)
between having the deeper information content inherent in longer shingles
and having enough of them.
The RMSE and MAE figures are again inflated
due to the presence of a small number of documents
that could not be dated accurately.\looseness=-1

\begin{figure}

\includegraphics{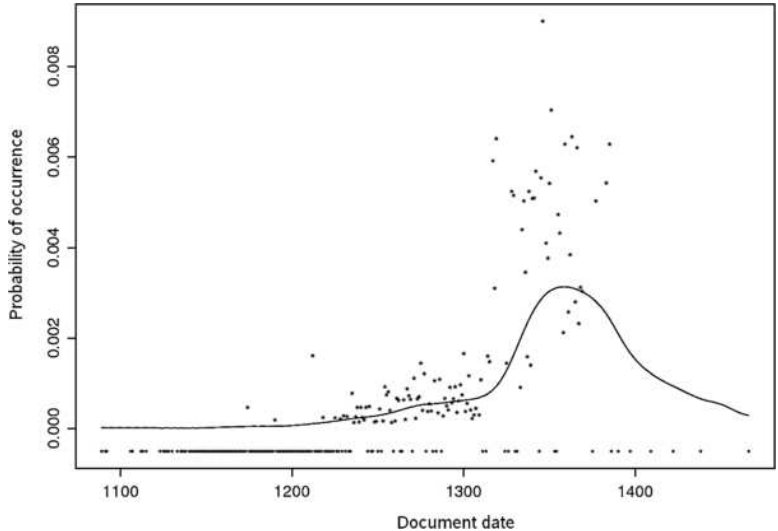}

\caption{Estimated probability function
$\hat\pi_s(t)$ for the shingle \textup{testimonium huic}
based on degrees of freedom $\nu= 3$ and bandwidth $h=12$.
The points are the relative frequencies for this shingle
at each date.}\label{Fig7p6a}\vspace*{-3pt}
\end{figure}

\begin{figure}

\includegraphics{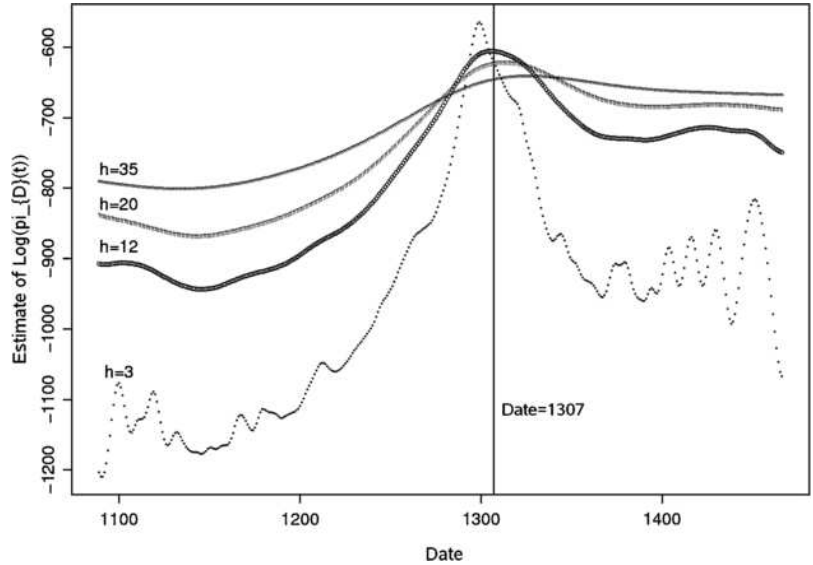}

\caption{Example of a prevalence function
$\hat\pi_{\cal D}(t)$, at four different bandwidths,
using a $t_3$ distribution kernel.
The true date for this document is 1299;
its maximum prevalence estimate is 1307.}\label{Fig6p13}
\end{figure}

\begin{figure}

\includegraphics{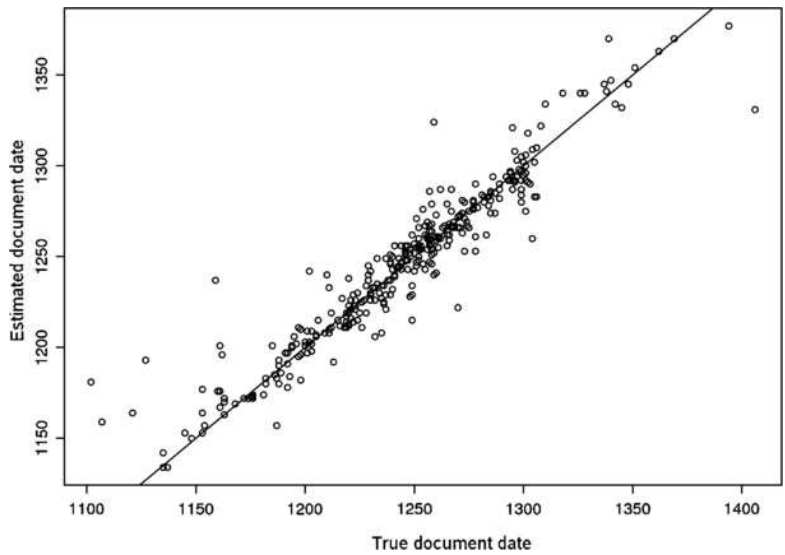}

\caption{Estimated versus actual dates
for the 326 documents in the test set ${\cal A}$,
using the maximum prevalence method with shingle length 2.
The solid line is ``$y=x$.''}\label{Fig6p4}
\end{figure}

Figures \ref{Fig7p6a}, \ref{Fig6p13} and \ref{Fig6p4}
exemplify the main components of the MP procedure.
Figure~\ref{Fig7p6a} shows an estimated probability function
$\hat\pi_s(t)$ for the 2-shingle \textit{testimonium huic}
(``in witness to which'')
based on a $t$-distribution kernel with bandwidth $h=12$
and degrees of freedom $\nu= 3$.
The points on this graph are the occurrence proportions
for this shingle over time,
and the concentration of points
%
at the bottom of the graph
correspond to years in which this shingle did not occur.
Figure \ref{Fig6p13} is a plot of the logarithm
of a typical prevalence curve $\hat\pi_{\cal D}(t)$,
based on shingle size $k=2$,
using four different bandwidths,
and a document ${\cal D}$ in the test set (consisting of 87 words)
whose true date is 1299.
The MP estimate for this document is 1307;
we note that (as was typically the case)
the resulting date estimate
is not unduly sensitive to the exact bandwidth chosen.
Figure \ref{Fig6p4} is a plot of the estimated versus the true dates
for the 326 documents in the test set using the M2 method.
Such edge bias as occurs could likely be reduced
by using the more computationally intensive
locally linear smoothing as in equations
(\ref{LocalLinearOne}) and (\ref{LocalLinearTwo}).\vadjust{\goodbreak}

We also attempted to combine the methods M1--M4
using a weighted average determined by minimizing
MSE (mean squared error) over the validation set
(subject to a constraint that the weights sum to 1).
The weights for the resulting method,
labeled M1234 in Table \ref{GelilaA},
were found to be 0.14, 0.64, 0.12 and 0.10.
The results for this method
were not much better than for M2 alone.

\begin{figure}

\includegraphics{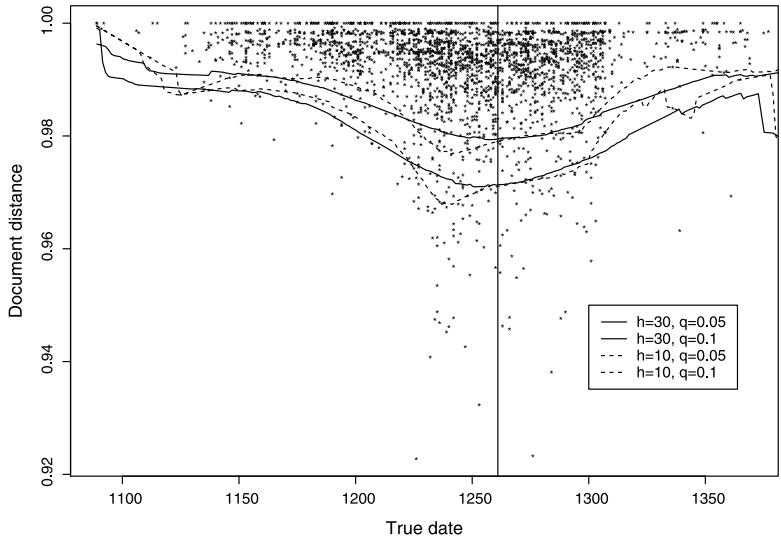}

\caption{Quantile regressions (QR) for
(lower) quantiles $q=0.1$ and $q=0.05$, using bandwidths $h=30$
(solid lines) and $h=10$ (dashed lines).
The points are distances from the document being dated
to documents in the test set,
plotted against the true dates of the test documents.
The vertical line is at the true date, 1261.}\label{FigQR}
\end{figure}

Our experiments with the QR method were less
successful than for the kNN and MP methods.
While the QR method did generally provide
meaningful estimates,
error variation was higher than for kNN or MP,
particularly for documents whose dates were in
the upper or lower date ranges
where test data was relatively sparse.
Figure \ref{FigQR} provides an illustration of the QR method
using a document ${\cal D}$ consisting of 336 words
whose true date is 1261
and a test set of 2608 documents~${\cal D}_i$.
In this plot of the distances $\operatorname{Dist}({\cal D},{\cal D}_i)$
versus the dates $t({\cal D}_i)$,
four quantile regression curves are drawn.
The two solid lines correspond to bandwidth
$h=30$, and the (lower) quantiles $q=0.1$ and $q=0.05$,
and lead to date estimates of 1256 and 1252, respectively;
the two dashed lines correspond to bandwidth
$h=10$, and (lower) quantiles $q=0.1$ and $q=0.05$,
and give date estimates 1240 and 1241.
Note that this plot is truncated at the far right
where the number of training documents
is too small to permit estimation
of the quantile curves at all reliably.

In a final series of experiments,
we attempted to combine the results of the kNN and MP methods.
For example, linearly combining M2 and kNN12
over the validation set using an RMSE criterion,
the optimal weights were found to be 0.83 and 0.17,
and the RMSE over the test set dropped slightly to 13.5 years.
The other performance measures, however,
were not significantly changed.\vspace*{-3pt}

\section{Discussion}
\label{SectionDiscussion}

The problem which motivated this work
leads to interesting technical questions
and novel techniques, linking statistical methods
to work associated with information retrieval.
Automated (i.e., computerized) calendaring
and temporal sequencing of text-based documents
are known to be difficult problems.
In the case of the DEEDS charters, however,
two features allow for progress to be made.
First, we have available a large (and increasing) training set
of documents whose dates are accurately known.
And second, the documents in question all have
relatively similar formulaic structure.

We remark that the methods we have described
can be applied to any collection of documents
and have potential applications broader
than the one which motivated this study.
For instance, as indicated in FHTG (\citeyear{Feuetal05}),
when suitable training data is available
kNN-based methods can be adapted to detect
other types of missing attributes, such as authorship,
potentially providing a methodology complementary
to that of \citet{MosWal63}.
Another potential application is in the detection of forgeries,
a problem related to that of establishing chronology
in that a common purpose of forgery is to alter past intent.
It is known that the number of forged
English medieval charters is not small.
One difficulty of this task, however,
is the fact that multiple and legitimate rewritings
of documents have been made
by scribes who may have modernized or slightly altered
the language of the documents being transcribed.
We also hope that the methods proposed here
may help determine more precise chronologies
in other contexts as well.

Of the methods investigated,
we found that the MP method performed best.
This appears to be due to its more detailed sensitivity
to the behavior of individual shingles over time.
For example, the MP method was more effective
in discounting very commonly occurring shingles,
since their occurrence probabilities were
relatively more constant over time.
In our numerical work, we also encountered two
somewhat surprising results.
The first is that of the shingle sizes we worked with,
shingles of size 1 resulted in estimates
not unduly far from the best results;
shingles of size 2 were better, but not by a large margin.
The second is that (to within the scale of our experiments)
combining multiple shingle sizes and combining methods
did not lead to striking improvements.
Taken together, these observations appear to suggest
that, for determining chronology, ``single words suffice.''

We are, however, not convinced that this observation
will be sustained by further work.
As the size of the DEEDS data set grows
and as our computing resources increase,
it will become possible to carry out estimation
using larger\vadjust{\goodbreak} training sets, using additional methods of estimation
and using more distances.
The situation is analogous to that encountered
in the collaborative filtering problem of the Netflix contest
where a blend of no fewer than 800 methods and variations
was needed by the winning team.
[See, e.g., \citet{FeuHeKha}.]
Thus, with more data, we expect further progress
to be possible via ensemble-type methods
and by blending methods differently
across strata of the data;
see, for example, \citet{HasTibFri09}, Chapter 16.
Further, with additional data, it will become feasible
to carry out optimization by referring undated documents
to other documents of their specific type only
(i.e., grant, lease, agreement, etc.),
and thus to tune the estimation procedures according to document type.
While further accuracy thus surely seems possible,
there must also be some practical limit to what can be achieved
via purely automated means,
particularly because any document to be dated is of finite length,
and therefore carries only a limited amount of ``information''
regardless of the amount of training data available.
While accuracies so far attained suffice to make
a material difference to historians studying that era,
the ultimate goal of the DEEDS project is to try to attain
an accuracy of about $\pm3$ years of error 95\% of the time.

We also expect that further progress could be made
on the definition of distances between documents.
One observation we offer
is that such distances should not be regarded as absolute,
but rather as relative to a particular collection of documents.
In this regard, the \textit{Multiplicador Total} method
of R. Fiallos seems particularly suggestive.
A highly effective distance between pairs of documents
should take into account all matching patterns between them,
as well as the lengths, lifetimes, currencies
and other relevant features
that these matching patterns possess within
the context of the \textit{whole} document collection.
Related to this is the degree of informativeness of shingles.
For example, \citet{Luh58} suggests that shingles which occur
neither too frequently nor too rarely will tend to be
the most informative.
As we had mentioned, our MP method does tend to discount
the very frequently occurring shingles,
but it does not discount the very rare ones.


The history of the DEEDS project is not yet fully written and
there is no doubt other techniques for the
calendaring problem will be explored.
For instance, in ongoing work, we are exploring ways
in which collections of documents can be
correctly \textit{sequenced} in time (to within time-reversal),
without regard to any of the dates associated with them.
We are also exploring ways in which methods such as
neural networks and support vector machines
might be applied to such calendaring problems.

Remarkably, during the time this work was being carried out, a medieval
English charter was discovered in a forgotten drawer of a library at
Brock University (near Niagara Falls), a discovery which resulted in a
certain amount of local media fanfare. This document records a land
grant from a certain Robert of Clopton to his son William. Attempts by
historians using paleography (analysis of handwriting), content and
other means initially attributed this document to the 14th century, and
subsequently to the 13th century. More careful work by Robin
Sutherland-Harris (a Ph.D. student of Medieval Studies at the
University of Toronto), based on the Patent Rolls (administrative
orders of the king) and the eyre records (records of the itinerant
courts), suggests a date range of 1235--1245, and perhaps, more
precisely, 1238--1242. These estimates are believed to be reliable; a
comparison document---believed to belong to the same time period---was
also found and was dated 1239. We dated this charter via maximum
prevalence (the most reliable among the methods we have discussed)
using our training set of 2608 documents; the date estimate we obtained
was 1246.

\section*{Acknowledgments}

It is a pleasure to acknowledge the generous assistance and substantial
contributions to this project by Rodolfo Fiallos. Thanks also to Robin
Sutherland-Harris for assistance with the Brock document. We also wish
to thank David Andrews, Michael Evans, Ben Kedem, Peter Hall, Keith
Knight, Radford Neil and Nancy Reid for their interest in this project
and for the benefit of many valuable discussions. We also thank the
referees for their careful reading and for suggestions which have
helped to improve the paper.



\printaddresses

\end{document}